\RequirePackage[mathlines]{lineno}
\documentclass[prl,twocolumn,showpacs,amsmath,amssymb]{revtex4-1}
\usepackage{overpic,graphicx}
\usepackage{dcolumn}
\usepackage{bm}
\usepackage{rotating}
\usepackage{subfigure}
\usepackage{color}
\usepackage{multirow}
\usepackage{tikz}

\begin{document}

\title{\bf \boldmath
First Experimental Study of the Purely Leptonic Decay $D_s^{*+}\to e^+\nu_e$
}

\author{
M.~Ablikim$^{1}$, M.~N.~Achasov$^{13,b}$, P.~Adlarson$^{75}$, R.~Aliberti$^{36}$, A.~Amoroso$^{74A,74C}$, M.~R.~An$^{40}$, Q.~An$^{71,58}$, Y.~Bai$^{57}$, O.~Bakina$^{37}$, I.~Balossino$^{30A}$
% ...
\\
\vspace{0.2cm}
(BESIII Collaboration)\\
\vspace{0.2cm} {\it
$^{1}$ Institute of High Energy Physics, Beijing 100049, People's Republic of China\\
% ...
$^{13}$ G.I. Budker Institute of Nuclear Physics SB RAS (BINP), Novosibirsk 630090, Russia\\
% ...
$^{30}$ INFN Sezione di Ferrara, (A)INFN Sezione di Ferrara, I-44122, Ferrara, Italy; (B)University of Ferrara,  I-44122, Ferrara, Italy\\
% ...
$^{36}$ Johannes Gutenberg University of Mainz, Johann-Joachim-Becher-Weg 45, D-55099 Mainz, Germany\\
$^{37}$ Joint Institute for Nuclear Research, 141980 Dubna, Moscow region, Russia\\
% ...
$^{40}$ Liaoning Normal University, Dalian 116029, People's Republic of China\\
% ...
$^{57}$ Southeast University, Nanjing 211100, People's Republic of China\\
$^{58}$ State Key Laboratory of Particle Detection and Electronics, Beijing 100049, Hefei 230026, People's Republic of China\\
% ...
$^{71}$ University of Science and Technology of China, Hefei 230026, People's Republic of China\\
% ...
$^{74}$ University of Turin and INFN, (A)University of Turin, I-10125, Turin, Italy; (B)University of Eastern Piedmont, I-15121, Alessandria, Italy; (C)INFN, I-10125, Turin, Italy\\
$^{75}$ Uppsala University, Box 516, SE-75120 Uppsala, Sweden\\
% ...
$^{b}$ Also at the Novosibirsk State University, Novosibirsk, 630090, Russia\\
% ...
}
}

\begin{abstract}
Using $7.33~\mathrm{fb}^{-1}$ of $e^+e^-$ collision data taken with the BESIII detector at the BEPCII collider, we report the first experimental study of the purely leptonic decay $D_s^{*+}\to e^+\nu_e$. A signal for the decay $D_s^{*+}\to e^+\nu_e$ is observed with a statistical significance of $2.9\sigma$. The branching fraction of ${D_s^{*+}\to e^+\nu_e}$ is measured to be $(2.1{^{+1.2}_{-0.9}}_{\rm stat.}\pm0.2_{\rm syst.})\times 10^{-5}$, corresponding to an upper limit of $4.0\times10^{-5}$ at the 90\% confidence level. Taking the total width of the $D_s^{*+}$~(($0.070\pm0.028$)\,keV) predicted by lattice quantum chromodynamics as input, the decay constant of the $D^{*+}_s$ is determined to be $f_{D_s^{*+}}=(213.6{^{+61.0}_{-45.8}}_{\rm stat.}\pm43.9_{\rm syst.})$\,MeV, corresponding to an upper limit of 353.8\,MeV at the 90\% confidence level.
\end{abstract}

\maketitle

\oddsidemargin  -0.2cm
\evensidemargin -0.2cm 

Purely leptonic decays of charmed-strange mesons, $D^{(*)+}_s\to \ell^+\nu_\ell$ ($\ell=e$, $\mu$ or $\tau$), offer the simplest and best-understood probes of the $c \to s$ quark transition~\cite{review}. The effects of the strong interaction can be parameterized in terms of the $D^{(*)+}_s$ decay constants~($f_{D^{(*)+}_s}$). Experimental studies of these decays are crucial to test lattice quantum chromodynamics~(LQCD) calculations of $f_{D^{(*)+}_s}$~\cite{Dslv1,theory0,theory1} and the unitarity of the quark-mixing matrix~\cite{Vcs1,Vcs2}. In addition, the branching fractions (BFs) of $D^{(*)+}_s\to \ell^+\nu_\ell$ for different families of leptons are important to test lepton flavor universality in the charm sector. Intense experimental investigations~\cite{DsLQCD1,DsLQCD2,DsLQCD3,dsmuv,tauevv,taupiv} of the (ground-state) pseudo-scalar mesons, e.g.~$D_s^+$, have allowed for precision tests of the standard model~(SM). However, for (excited-state) vector mesons, e.g.~$D_s^{*+}$, there have been relatively few theoretical studies and no experimental study of their weak decays has yet been reported.\par

Reference~\cite{theory0} states that the $D_s^{*+}\to e^+\nu_e$ decay may be the most promising channel to observe the weak decay of a charmed vector meson. In the SM, the decay width of $D_s^{*+}\to \ell^+\nu_{\ell}$ can be written as
~\cite{theory1}
\begin{linenomath*}
\small
\begin{multline}
  \Gamma(D_s^{*+}\to \ell^+\nu_{\ell})=\\
  \frac{G_F^2}{12\pi}\vert V_{cs}\vert^2f_{D_s^{*+}}^2m_{D_s^{*+}}^3\left (1-\frac{m_{\ell^+}^2}{m_{D_s^{*+}}^2} \right )^2 \left (1+\frac{m_{\ell^+}^2}{2m_{D_s^{*+}}^2}\right ),
  \label{eq:Gamma}
\end{multline}
\end{linenomath*}
\\
where
$G_F$ is the Fermi coupling constant,
$\vert V_{cs}\vert$ is the $c\to s$ Cabibbo-Kobayashi-Maskawa matrix element,
$m_{\ell^+}$ is the lepton mass,
and $m_{D_s^{*+}}$ is the $D_s^{*+}$ mass.
Its decay BF is predicted to be up to $10^{-5}$ according to Refs.~\cite{theory0,theory1,theory2}. The decay constant $f_{D_s^{*+}}$ has been calculated via the nonrelativistic quark model~\cite{fdst1,fdst2,fdst3}, the relativistic quark model~\cite{fdst4,fdst5,fdst6}, the light-front quark model~\cite{fdst7,fdst8,fdst9}, QCD sum rules~\cite{fdst10,fdst11,fdst12}, and LQCD~\cite{theory1,fdst13,fdst14,fdst15,fdst16,fdst17} with predicted values varying from 212~\cite{fdst3} to 447~\cite{fdst1} MeV. Measurement of the $D_s^{*+}\to e^+\nu_e$ decay is crucial to test these theoretical calculations. The obtained results could also be used to indirectly constrain the unknown total decay width of the $D_s^{*+}$, thereby clarifying the large differences in various theoretical predictions of the electromagnetic and strong couplings~\cite{theory0,coupling1,coupling2,coupling3,coupling4,coupling5}.

This letter presents the first experimental search for the decay $D_s^{*+}\to e^+\nu_e$ by using $7.33~\mathrm{fb}^{-1}$~\cite{Datasamples} 
of $e^+e^-$ collision data collected by the BESIII detector at the BEPCII collider at the center-of-mass 
energies $E_{\rm cm}=4.128$, 4.157, 4.178, 4.189, 4.199, 4.209, 4.219, and 4.226\,GeV. 
At these energies, the $D_s^{*+}$ mesons are produced mainly through the process $e^+e^-\to D_s^-D_s^{*+}+c.c$. In an event where a $D_s^-$ meson is fully reconstructed and comes directly from the $e^+e^-$ collision~(called a single-tag (ST) $D_s^-$ meson), the $D_s^{*+}$ meson decaying to $e^+\nu_e$ can be searched for in the recoiling system. Surviving events are called double-tag (DT) events. Throughout this letter, charge-conjugate modes are always implied.\par

A description of the design and performance of the BESIII detector is given in Ref.~\cite{bes3des}. The endcap time-of-flight~(TOF) system was upgraded with multi-gap resistive plate chamber technology and now has a time resolution of 60~ps~\cite{mrpc1,mrpc2}. Monte Carlo~(MC) events are generated with a {\sc geant}4-based~\cite{geant4} detector simulation software package~\cite{boost}, which includes both the geometrical description and the response of the detector~\cite{detector}. 
Inclusive MC samples are produced at the corresponding center-of-mass energies and include all open charm processes, 
initial state radiation~(ISR) production of the $\psi(3770)$, $\psi(3686)$ and $J/\psi$, and $q\bar{q}\,(q=u,d,s)$ continuum processes, 
along with Bhabha scattering, $\mu^+\mu^-$, $\tau^+\tau^-$, and $\gamma\gamma$ events. 
The open charm processes are generated using {\sc conexc}~\cite{conexc}. 
The effects of ISR~\cite{isr} and final state radiation~(FSR)~\cite{photons} are considered. 
The decay modes with known BFs are generated using {\sc evtgen}~\cite{evtgen1, evtgen2} 
and the other modes are generated using {\sc lundcharm}~\cite{lundcharm}.
Also, a dedicated ``signal MC'' sample of $e^+e^-\to D_s^-D_s^{*+}+c.c.$ was generated in which $D_s^-$ is allowed 
to decay via all possible channels and the $D_s^{*+}$ decays to $e^+\nu_e$.\par

The ST $D_s^-$ mesons are reconstructed from 16 hadronic decay modes,
$D_s^-\to K^+K^-\pi^-$,
$K^+K^-\pi^-\pi^0$,
$\pi^+\pi^-\pi^-$,
$K_S^0K^-$,
$K_S^0K^-\pi^0$,
$K^-\pi^+\pi^-$,
$K_S^0K_S^0\pi^-$,
$K_S^0K^+\pi^-\pi^-$,
$K_S^0K^-\pi^+\pi^-$,
$\eta_{\gamma\gamma}\pi^-$,
$\eta_{\pi^+\pi^-\pi^0}\pi^-$,
$\eta'_{\pi^+\pi^-\eta_{\gamma\gamma}}\pi^-$,
$\eta'_{\gamma\rho^0}\pi^-$,
$\eta_{\gamma\gamma}\rho^-$,
$\eta_{\pi^+\pi^-\pi^0}\rho^-$, and
$\eta_{\gamma\gamma}\pi^+\pi^-\pi^-$,
where the subscripts of $\eta$~($\eta^{\prime}$) represent the decay modes used to reconstruct $\eta$~($\eta^{\prime}$). Throughout the text, $\rho$ denotes $\rho(770)$ and $\rho^{-/0}$ decays to $\pi^-\pi^{0/+}$.\par

All charged tracks except those from $K_S^0$ must originate from the interaction point with a distance of closest approach less than 1 cm in the transverse plane and less than 10 cm along the $z$ axis. The polar angle $\theta$ of each track defined with respect to the symmetry axis of the main drift chamber~(MDC) must satisfy $|\cos\theta|<0.93$. The combined information from the specific ionization energy loss (d$E$/d$x$) measured by the MDC, the TOF, and the electromagnetic calorimeter~(EMC) are used for particle identification~(PID) by forming likelihoods $\mathcal{L}_p~(p=K, \pi, e)$ for each particle $p$ hypothesis. Kaon~(pion) candidates are required to satisfy $\mathcal{L}_{K(\pi)}>\mathcal{L}_{\pi(K)}$ and $\mathcal{L}_{K(\pi)}>\mathcal{L}_{e}$.\par

To select $K_S^0$ candidates, pairs of oppositely charged tracks with distances of closest approach to the interaction point less than 20 cm along the $z$ axis are assigned as $\pi^+\pi^-$ without PID requirements. These $\pi^+\pi^-$ combinations are required to have an invariant mass within $\pm12$\,MeV$/c^{2}$ of the world average $K_S^0$ mass~\cite{PDG2022} and have a decay length greater than twice its resolution.\par

The $\pi^0$ and $\eta$ mesons are reconstructed from photon pairs. Photon candidates are selected from the shower clusters in the EMC. Each electromagnetic shower is required to start within [0, 700] ns from the event start time. The shower energy is required to be greater than 25\,(50)\,MeV in the barrel\,(endcap) region of the EMC~\cite{taulv}. The opening angle between the candidate shower 
and the nearest charged track extrapolated to the EMC is required to be greater than $10^{\circ}$. To form $\pi^0$ and $\eta$ candidates, the invariant masses of the selected photon pairs are required to be within the $M_{\gamma\gamma}$ intervals $(115,\,150)$ and $(500,\,570)$\,MeV$/c^{2}$, respectively. To improve momentum resolution, a kinematic fit is imposed on each chosen photon pair to constrain its invariant mass to the world average $\pi^{0}$ or $\eta$ mass~\cite{PDG2022}.\par

For the tag modes $D_s^-\to \eta_{\pi^0\pi^+\pi^-}\pi^-$ and $\eta_{\pi^0\pi^+\pi^-}\rho^-$, the $\pi^0\pi^+\pi^-$ combinations used to form $\eta$ candidates are required to be within the intervals $(530,\,570)~\mathrm{MeV}/c^2$. The $\eta^\prime$ mesons are reconstructed via two decay modes, $\eta\pi^+\pi^-$ and $\gamma\rho^0$, whose invariant masses are required to be within the intervals $(946,\,970)$ and $(940,\,976)~\mathrm{MeV}/c^2$, respectively. In addition, the energy of the $\gamma$ from $\eta'\to\gamma\rho^0$ decays must be greater than 100\,MeV. The $\rho^0$ and $\rho^-$ candidates are reconstructed from the $\pi^+\pi^-$ and $\pi^-\pi^0$ combinations with invariant masses within the interval $(570,\,970)~\mathrm{MeV}/c^2$.\par

To remove soft pions originating from $D^*$ transitions, the momentum of pions directly from the ST $D_s^-$ are required to be greater than 100\,MeV/$c$. For the tag mode $D_s^-\to \pi^+\pi^-\pi^-$~($D_s^-\to K^-\pi^+\pi^-$), the contribution of the peaking background from the decay $D_s^-\to K^0_S\pi^-$~($D_s^-\to K^0_SK^-$) is rejected by requiring all $\pi^+\pi^-$ combinations to be outside the mass window of $(468,\,528)~\mathrm{MeV}/c^2$.\par

\begin{figure}
  \centering
  \begin{tikzpicture}
  \node [ above right, inner sep=0] (image) at (0,0) {\includegraphics[keepaspectratio=true,width=0.50\textwidth,angle=0]{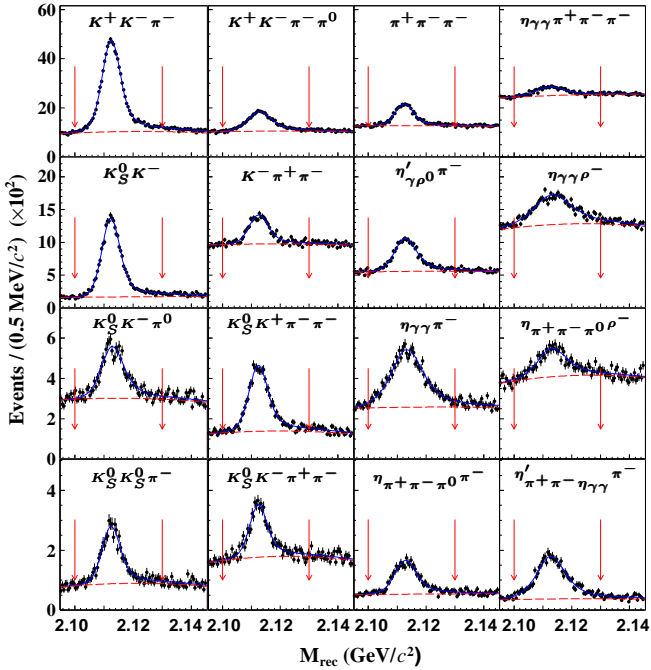}};
  \node[align=left] at (1.75,8.70) {\tiny $\pmb{K^+K^-\pi^-}$};
  \node[align=left] at (3.75,8.70) {\tiny $\pmb{K^+K^-\pi^-\pi^0}$};
  \node[align=left] at (5.70,8.70) {\tiny $\pmb{\pi^+\pi^-\pi^-}$};
  \node[align=left] at (7.65,8.70) {\tiny $\pmb{\eta_{\gamma\gamma}\pi^+\pi^-\pi^-}$};
  \node[align=left] at (1.75,6.65) {\tiny $\pmb{K_S^0K^-}$};
  \node[align=left] at (3.75,6.65) {\tiny $\pmb{K^-\pi^+\pi^-}$};
  \node[align=left] at (5.70,6.65) {\tiny $\pmb{\eta^{\prime}_{\gamma\rho^0}\pi^-}$};
  \node[align=left] at (7.65,6.65) {\tiny $\pmb{\eta_{\gamma\gamma}\rho^-}$};
  \node[align=left] at (1.75,4.65) {\tiny $\pmb{K_S^0K^-\pi^0}$};
  \node[align=left] at (3.75,4.65) {\tiny $\pmb{K_S^0K^+\pi^-\pi^-}$};
  \node[align=left] at (5.70,4.65) {\tiny $\pmb{\eta_{\gamma\gamma}\pi^-}$};
  \node[align=left] at (7.65,4.65) {\tiny $\pmb{\eta_{\pi^+\pi^-\pi^0}\rho^-}$};
  \node[align=left] at (1.75,2.60) {\tiny $\pmb{K_S^0K_S^0\pi^-}$};
  \node[align=left] at (3.75,2.60) {\tiny $\pmb{K_S^0K^-\pi^+\pi^-}$};
  \node[align=left] at (5.70,2.60) {\tiny $\pmb{\eta_{\pi^+\pi^-\pi^0}\pi^-}$};
  \node[align=left] at (7.65,2.60) {\tiny $\pmb{\eta^{\prime}_{\pi^+\pi^-\eta_{\gamma\gamma}}\pi^-}$};
  \end{tikzpicture}
  \caption{\footnotesize
  Fits to the $M_{\rm rec}$ distributions of the ST candidates at $E_{\rm cm}=4.178$\,GeV.
  Points with error bars are data.
  Blue solid curves are the fit results.
  Red dashed curves are the fitted backgrounds.
  Pairs of red arrows denote the signal regions.
  }
  \label{fig:STfit}
\end{figure}

In each event, we only keep the ST $D_s^-$ candidate with mass~($M_{D_s^-}$) closest to the
world average value~\cite{PDG2022} per tag mode per charge. Non-$D_s^-D_s^{*+}$ events are further suppressed by requiring that $M_{D_s^-}$ agrees with the world average value within 3$\sigma$~\cite{dsmuv, tauevv,bes3_etaev} of each tag mode experimental resolution. The $M_{D_s^-}$ requirements are listed in the second column of Table~\ref{tab:eff1}. The recoiling mass of the ST $D_s^-$ is defined as
\begin{equation}
  M_{\rm rec}\equiv \sqrt{ \left (E_{\rm cm}-\sqrt{|\vec p_{D_s^-}|^2c^2+m^2_{D_s^-}c^4} \right )^2/c^4-|\vec p_{D_s^-}|^2/c^2},
\end{equation}
where $(\vec p_{\rm cm}, E_{\rm cm})$ is the four momentum of the $e^+e^-$ system and $\vec p_{D_s^-}$ is the measured momentum of the ST $D_s^-$ candidate. The $m_{D_s^-}$ is fixed at the world average $D_s^-$ mass. The $M_{\rm rec}$ distribution of the direct $D_s^-$ from a $e^+e^-\to D_s^-D_s^{*+}$ pair will form a peak around the world average $D_s^{*+}$ mass~\cite{PDG2022}, while other processes~(e.g. $e^+e^-\to \gamma D_s^+D_s^-$, $e^+e^-\to \pi^0 D_s^+D_s^-$) produce flat distributions. The ST yield for each tag mode is extracted from the fit to the corresponding $M_{\rm rec}$ spectrum. The signal is described by the MC-simulated shape convolved with a Gaussian function representing the resolution difference between data and MC simulation. The non-peaking background is modeled by a second- or third-order Chebychev polynomial function. 
The parametrization of the background shape is validated using the inclusive MC sample.
Events for the further analysis are selected within the 
$M_{\rm rec}$ signal regions optimized according to the figure-of-merit ($S/\sqrt{S+B}$) at each center-of-mass energy. 
Here, $S$ and $B$ denote the signal and background yields from the inclusive MC sample. 
The $E_{\rm cm}$-dependent $M_{\rm rec}$ regions are listed in the second column of Table~\ref{tab:eff2}. As an example, Fig.~\ref{fig:STfit} shows the fit results of the $M_{\rm rec}$ spectra for various tag modes in data at $E_{\rm cm}=4.178$\,GeV. The resulting ST yields ($N_{\rm ST}^{i~4.178}$) and the corresponding ST efficiencies ($\varepsilon_{\rm ST}^{i~4.178}$) are summarized in the third and fourth columns of Table~\ref{tab:eff1}, respectively. The results for $N^{ij}_{\rm ST}$ and $\varepsilon_{\rm ST}^{ij}$ at the other energy points are obtained similarly. The total ST yields $N^j_{\rm ST}$ at various energy points are summarized in the third column of Table~\ref{tab:eff2}. The $i$ denotes the 16 ST modes and the $j$ denotes different data samples.
\begin{table}[htbp]
  \centering\linespread{1.15}
  \caption{
  Requirements for $M_{D^-_s}$ and the obtained values of $N_{\rm ST}^{i~4.178}$, $\varepsilon_{\rm ST}^{i~4.178}$, and $\varepsilon_{\rm DT}^{i~4.178}$ in the $i$-th tag mode at $E_{\rm cm}=4.178$\,GeV, where the uncertainties are statistical only. The differences among the ratios of $\varepsilon_{\rm DT}^{i~4.178}$ over $\varepsilon_{\rm ST}^{i~4.178}$ for various modes are mainly due to the $M_{\rm rec}$ and other signal side requirements.}
  \small
  \label{tab:eff1}
  \begin{tabular}{l|cr@{}lr@{}lr@{}l}\hline\hline
  \multirow{2}{*}{Tag mode}&$M_{D_s^-}$&\multicolumn{2}{c}{$N_{\rm ST}^{i~4.178}$}&\multicolumn{2}{c}{$\varepsilon_{\rm ST}^{i~4.178}$}&\multicolumn{2}{c}{$\varepsilon_{\rm DT}^{i~4.178}$} \\
  &(GeV/$c^2$)&\multicolumn{2}{c}{($\times 10^3$)}&\multicolumn{2}{c}{(\%)}&\multicolumn{2}{c}{(\%)} \\
  \hline
  $K^+K^-\pi^-$&[1.950, 1.986]&56.3&$\pm$0.4&34.67&$\pm$0.04&24.46&$\pm$0.11\\
  $K^+K^-\pi^-\pi^0$&[1.947, 1.982]&18.3&$\pm$0.5&10.62&$\pm$0.03&8.57&$\pm$0.07\\
  $\pi^+\pi^-\pi^-$&[1.952, 1.984]&15.8&$\pm$0.4&46.97&$\pm$0.14&35.19&$\pm$0.13\\
  $K_S^0K^-$&[1.948, 1.991]&14.0&$\pm$0.2&43.41&$\pm$0.08&31.71&$\pm$0.13\\
  $K_S^0K^-\pi^0$&[1.946, 1.987]&4.9&$\pm$0.2&15.91&$\pm$0.09&13.44&$\pm$0.08\\
  $K^-\pi^+\pi^-$&[1.953, 1.983]&8.2&$\pm$0.4&40.16&$\pm$0.22&29.45&$\pm$0.12\\
  $K_S^0K_S^0\pi^-$&[1.951, 1.986]&2.3&$\pm$0.1&20.82&$\pm$0.12&14.85&$\pm$0.09\\
  $K_S^0K^+\pi^-\pi^-$&[1.953, 1.983]&6.1&$\pm$0.2&18.11&$\pm$0.06&13.07&$\pm$0.08\\
  $K_S^0K^-\pi^+\pi^-$&[1.958, 1.980]&3.1&$\pm$0.2&16.12&$\pm$0.10&12.69&$\pm$0.08\\
  $\eta_{\gamma\gamma}\pi^-$&[1.930, 2.000]&8.4&$\pm$0.3&42.61&$\pm$0.18&34.95&$\pm$0.13\\
  $\eta_{\pi^+\pi^-\pi^0}\pi^-$&[1.941, 1.990]&2.4&$\pm$0.1&20.76&$\pm$0.13&16.87&$\pm$0.09\\
  $\eta'_{\pi^+\pi^-\eta_{\gamma\gamma}}\pi^-$&[1.940, 1.996]&3.9&$\pm$0.1&20.48&$\pm$0.10&16.14&$\pm$0.09\\
  $\eta'_{\gamma\rho^0}\pi^-$&[1.938, 1.992]&10.7&$\pm$0.3&29.21&$\pm$0.12&23.23&$\pm$0.11\\
  $\eta_{\gamma\gamma}\rho^-$&[1.920, 2.006]&15.3&$\pm$0.6&16.95&$\pm$0.09&16.51&$\pm$0.09\\
  $\eta_{\pi^+\pi^-\pi^0}\rho^-$&[1.927, 1.997]&3.9&$\pm$0.3&7.52&$\pm$0.07&7.46&$\pm$0.06\\
  $\eta_{\gamma\gamma}\pi^+\pi^-\pi^-$&[1.946, 1.990]&9.6&$\pm$0.7&21.37&$\pm$0.22&19.99&$\pm$0.10\\
  \hline\hline
  \end{tabular}
\end{table}

\begin{table}[htbp]\centering\linespread{1.15}
  \caption{
    The total ST yields ($N_{\rm ST}^j$) and the averaged signal efficiencies ($\varepsilon^j_{e^+\nu_e}$) for data samples at different energy points, where the uncertainties are statistical only. The differences among the signal efficiencies for different data samples are mainly due to the $M_{\rm rec}$ requirement.}
  \small
  \label{tab:eff2}
  \begin{tabular}{l|cr@{}lr@{}l}\hline\hline
  $E_{\rm cm}$~(GeV)&$M_{\rm rec}$~(GeV/$c^2$)&\multicolumn{2}{c}{$N_{\rm ST}^j$~($\times 10^3$)}&\multicolumn{2}{c}{$\varepsilon^j_{e^+\nu_e}$~(\%)}\\
  \hline
  4.128&$[2.105, 2.123]$&14.8&$\pm$0.3&81.31&$\pm$0.35\\
  4.157&$[2.103, 2.127]$&22.0&$\pm$0.5&78.58&$\pm$0.29\\
  4.178&$[2.100, 2.130]$&183.1&$\pm$1.4&77.01&$\pm$0.16\\
  4.189&$[2.100, 2.136]$&32.5&$\pm$0.7&77.25&$\pm$0.28\\
  4.199&$[2.099, 2.140]$&30.1&$\pm$0.9&76.61&$\pm$0.26\\
  4.209&$[2.098, 2.145]$&30.9&$\pm$0.6&74.22&$\pm$0.24\\
  4.219&$[2.098, 2.154]$&26.0&$\pm$0.5&74.49&$\pm$0.28\\
  4.226&$[2.098, 2.167]$&41.1&$\pm$0.7&74.94&$\pm$0.24\\
  \hline\hline
  \end{tabular}
\end{table}

\begin{figure}[htbp]
\centering
\includegraphics[width=0.50\textwidth]{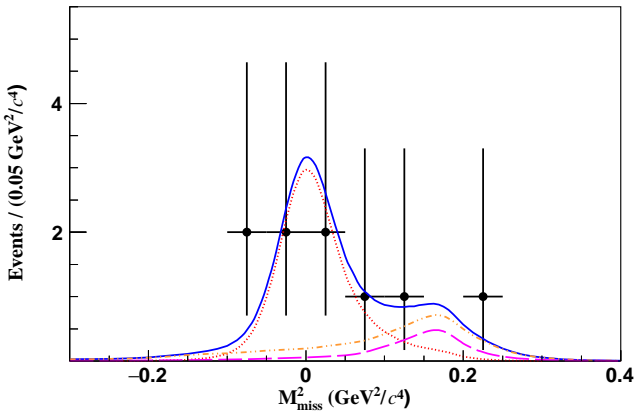}
\caption{\footnotesize
Fit to the $M_{\rm miss}^2$ distribution of the $D_s^{*+}\to e^+\nu_e$ candidates.
Points with error bars are data.
The blue solid curve is the fit result.
The red dotted curve is the signal.
The brown dotted-dashed curve is the total fitted background.
The pink dashed curve denotes background from the decay $D_s^{+}\to \tau^+\nu_\tau$ with $\tau^+\to e^+\nu_e\bar{\nu}_{\tau}$.
}
\label{fig:DTfit}
\end{figure}

On the recoiling side of the ST $D_s^-$ meson, we require that there is only one residual charged track to be identified as an $e^+$. The $e^+$ candidate is required to satisfy $\mathcal{L}_e>0.8\times(\mathcal{L}_e+\mathcal{L}_{\pi}+\mathcal{L}_K)$ and $\mathcal{L}_e>0.001$. The energy loss of the $e^+$ candidate due to bremsstrahlung is partially recovered by adding the energies of the EMC showers that are within $10^{\circ}$ of the $e^+$ direction and not matched to other particles. 
The signal is separated from combinatorial backgrounds using the square of the missing mass defined as
\begin{equation}
  M_{\rm miss}^2\equiv |E_{\rm cm}-\sum\nolimits_kE_k|^2/c^4 - |\sum\nolimits_k\vec{p}_k|^2/c^2,
\end{equation}
where $E_k$ and $\vec{p}_k$ are the energy and momentum of particle $k$ in the center-of-mass frame, and the sum runs over the ST $D_s^-$ and the $e^+$ of the signal side. The measured energy of the $e^+$ in the rest system of the $D_s^{*+}$ is required to be greater 
than 1.01\,GeV based on optimization. 
To suppress backgrounds from hadrons and muons the $e^+$ candidate is required 
to have the ratio of the energy deposited in the EMC over the momentum ($E/p$) within the range  [0.8, 1.2].
To suppress background events with extra photon(s), the maximum energy of the unused showers in the DT selection~($E_{\mathrm{extra}~\gamma}^{\rm max}$) is required to be less than 300\,MeV. Figure~\ref{fig:DTfit} shows the $M_{\rm miss}^2$ distribution for the accepted DT candidate events in data.\par
To extract the DT yield, an un-binned maximum extended likelihood fit is performed to the resulting $M_{\rm miss}^2$ distribution. 
In the fit, the signal and background shapes are obtained from the signal and inclusive MC samples, respectively. 
The background from the decay $D_s^{+}\to \tau^+\nu_\tau$ with $\tau^+\to e^+\nu_e\bar{\nu}_{\tau}$ tends to form a peak around 0.17\,GeV$^2$ 
while other backgrounds are flat. We combine all backgrounds from the inclusive MC into one background shape. 
The event yields from the signal and the combined background are free parameters of the fit. The results of the fit 
to the $M_{\rm miss}^2$ distribution are shown in Fig.~\ref{fig:DTfit}. The obtained signal yield of $D_s^{*+}\to e^+\nu_e$ 
is $N_{\rm DT}=6.2^{+3.4}_{-2.7}$, where the uncertainty is statistical only. 
The statistical significance of the signal is 2.9$\sigma$, which is estimated by comparing the likelihoods 
with and without the signal component in the fit and taking into account the number of degrees of freedom.\par

The efficiencies for reconstructing the DT candidate events are determined with the signal MC sample. 
The DT efficiencies ($\varepsilon^{i~4.178}_{\rm DT}$) obtained at $E_{\rm cm}=4.178$\,GeV are summarized in the fifth column of Table~\ref{tab:eff1}. Dividing the DT efficiencies by the ST efficiencies yields the corresponding efficiencies for detecting $D^{*+}_s\to e^+\nu_e$. The obtained $\varepsilon^{j}_{e^+\nu_e}$ at various energy points are summarized in the fourth column of Table~\ref{tab:eff2}. 
The effective signal efficiency for finding $e^+\nu_e$, weighted by the ST yields for different tag modes and energy points, is obtained to be $\bar{\varepsilon}_{e^+\nu_e}=(76.63\pm0.09)\%$. The BF of $D_s^{*+}\to e^+\nu_e$ is determined by
\begin{equation}
\mathcal{B}({D_s^{*+}\to e^+\nu_e})=\frac{N_{\rm DT}}{\bar{\varepsilon}_{e^+\nu_e}\sum_jN_{\rm ST}^j},
\end{equation}
to be $(2.1{^{+1.2}_{-0.9}}_{\rm stat.}\pm0.2_{\rm syst.})\times10^{-5}$. The statistical uncertainty is from $N_{\rm DT}$ and the systematic uncertainty is discussed later.

Because of limited statistics, we also set an upper limit on the BF of $D_s^{*+}\to e^+\nu_e$ by following Refs.~\cite{upperlimit1,upperlimit2,upperlimit3} after incorporating the systematic uncertainty via a likelihood scan method. The upper limit of the BF of $D_s^{*+}\to e^+\nu_e$ is obtained to be $4.0\times10^{-5}$ at the 90\% confidence level~(C.L.).\par

The total systematic uncertainty of the BF measurement is determined to be 10.8\%.
It is obtained by adding in quadrature the individual contributions described below.

The $e^+$ tracking and PID efficiencies~(including the $E/p$ requirement) are studied with radiative Bhabha scattering events. The efficiency differences between data and MC simulation, 0.5\% and 1.0\%, are assigned as the corresponding systematic uncertainties.
The uncertainty in the yield of ST $D_s^-$ mesons is assigned to be 0.9\% by examining the relative changes of the fit yields 
when varying the criteria of truth-matching for signal shape and the order of Chebychev function for background shape. 
The uncertainty due to the MC statistics is 0.1\%. The uncertainty due to the FSR effect is assigned to be 0.5\% by studying the radiative Bhabha scattering events~\cite{tauevv}.
The uncertainty due to different multiplicities of tag environments are assigned as 0.5\%~\cite{dsmuv}.\par
The efficiency for the requirements of $E_{\mathrm{extra}~\gamma}^{\rm max}$ and only one charged track in the signal side is studied with the hadronic DT samples. The systematic uncertainty is taken to be 0.5\% considering the efficiency differences between data and MC simulation. For the requirement of the $e^+$ energy in the rest system of $D_s^{*+}$, the systematic uncertainty is estimated by changing this requirement by $\pm10$\,MeV. The largest relative change of the BF, 9.5\%, is assigned as the corresponding systematic uncertainty.\par

The systematic uncertainty of the signal shape is estimated by using an alternative MC model and is found to be negligible.
The systematic uncertainty in the background shape is examined by using alternative background shapes obtained by varying the relative fractions of the different background components and shifting the input cross sections by $\pm 1\sigma$. The relative change of the BF, 4.8\%, is taken as the systematic uncertainty.\par

With Eq.~\ref{eq:Gamma} and the equation 1 of Ref.~\cite{dsmuv}, the total width of the $D^{*+}_s$ is expected to be $\Gamma^{\rm total}_{D^{*+}_s}=2.04\times10^{-3}\times(\frac{f_{D^{*+}_s}}{f_{D^{+}_s}})^2/{\mathcal B}(D^{*+}_s\to e^+\nu_e)$ eV, after combining the world average values of ${\mathcal B}(D^+_s\to \mu^+\nu_\mu)$, the lifetime of the $D^+_s$, $m_e$, $m_{D_s^{*+}}$, $m_\mu$, and $m_{D_s^{+}}$~\cite{PDG2022}.~Combining with $\frac{f_{D^{*+}_s}}{f_{D^{+}_s}}=1.12\pm0.01$ averaged from LQCD calculations~\cite{theory1,fdst13,fdst14,fdst15,fdst16,fdst17} and ${\mathcal B}(D^{*+}_s\to e^+\nu_e)$ obtained in this work gives $\Gamma^{\rm total}_{D^{*+}_s}=(121.9^{+69.6}_{-52.2}\pm11.8$)\,eV. It agrees with $(70\pm28)$\,eV predicted by LQCD~\cite{theory1} within $\pm 1\sigma$.\par

Combining our BF measurement with the world average values of $G_F$, $m_e$, $m_{D_s^{*+}}$~\cite{PDG2022}, and $\Gamma^{\rm total}_{D^{*+}_s}$ given by LQCD~\cite{theory1}, we obtain $f_{D_s^{*+}}|V_{cs}|=(207.9{^{+59.4}_{-44.6}}_{\rm stat.}\pm42.7_{\rm syst.})$\,MeV. Here the systematic uncertainty is mainly from the uncertainties in the measured ${\mathcal B}(D^{*+}_s\to e^+\nu_e)$~(10.8\%) and the LQCD predicted $\Gamma^{\rm total}_{D^{*+}_s}$~(40.0\%). Taking $|V_{cs}|=0.97349\pm0.00016$ from the SM global fit~\cite{PDG2022} as input, we determine $f_{D_s^{*+}}=(213.6{^{+61.0}_{-45.8}}_{\rm stat.}\pm43.9_{\rm syst.})$\,MeV, corresponding to an upper limit of 353.8\,MeV at the 90\% C.L..\par

\begin{figure}[htbp]
  \centering
  \includegraphics[width=0.50\textwidth]{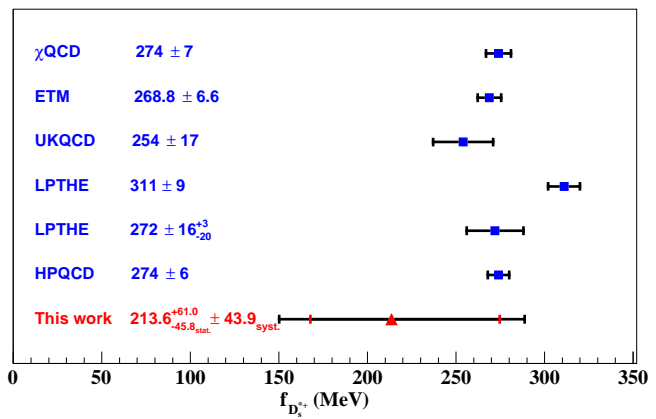}
  \caption{\footnotesize
  Comparison of $f_{D_s^{*+}}$ obtained in this work and LQCD calculations. For the result of this work, the shorter red error bar denotes the statistical uncertainly only while the longer black error bar combines both statistical and systematic uncertainties.
  }
  \label{fig:fdsST}
  \end{figure}

In summary, by analyzing $7.33~\mathrm{fb}^{-1}$ of $e^+e^-$ collision data collected at $E_{\rm cm}$ from 4.128 to 4.226\,GeV with the BESIII detector, we report the first experimental search for the purely leptonic decay $D_s^{*+}\to e^+\nu_e$. The BF of $D_s^{*+}\to e^+\nu_e$ is determined to be $(2.1{^{+1.2}_{-0.9}}_{\rm stat.}\pm0.2_{\rm syst.})\times 10^{-5}$. Our result indirectly constrains the upper limit on the total width $\Gamma^{\rm total}_{D^{*+}_s}$ from MeV~\cite{PDG2022} to keV level. Using the $\Gamma_{D^{*+}_s}$ predicted by LQCD and the $|V_{cs}|$ extracted by the global fit in the SM, we obtain the decay constant of $D_s^{*+}$. Figure~\ref{fig:fdsST} shows the comparison of the $f_{D_s^{*+}}$ obtained in this work and the LQCD calculations from HPQCD~\cite{theory1}, LPTHE~\cite{fdst13,fdst14}, UKQCD~\cite{fdst15}, ETM~\cite{fdst16} and $\chi$QCD~\cite{fdst17}. The obtained $f_{D_s^{*+}}$ offers the first experimental test on various theoretical calculations. This analysis opens an avenue to study the weak decays of charmed vector mesons in experiments. \par

\begin{acknowledgments}

The BESIII Collaboration thanks the staff of BEPCII and the IHEP computing center for their strong support. This work is supported in part by National Key R\&D Program of China under Contracts Nos. 2020YFA0406300, 2020YFA0406400; National Natural Science Foundation of China (NSFC) under Contracts Nos. 11635010, 11735014, 11835012, 11935015, 11935016, 11935018, 11961141012, 12022510, 12025502, 12035009, 12035013, 12061131003, 12192260, 12192261, 12192262, 12192263, 12192264, 12192265; the Chinese Academy of Sciences (CAS) Large-Scale Scientific Facility Program; the CAS Center for Excellence in Particle Physics (CCEPP); Joint Large-Scale Scientific Facility Funds of the NSFC and CAS under Contract No. U1832207; CAS Key Research Program of Frontier Sciences under Contracts Nos. QYZDJ-SSW-SLH003, QYZDJ-SSW-SLH040; 100 Talents Program of CAS; The Institute of Nuclear and Particle Physics (INPAC) and Shanghai Key Laboratory for Particle Physics and Cosmology; ERC under Contract No. 758462; European Union's Horizon 2020 research and innovation programme under Marie Sklodowska-Curie grant agreement under Contract No. 894790; German Research Foundation DFG under Contracts Nos. 443159800, 455635585, Collaborative Research Center CRC 1044, FOR5327, GRK 2149; Istituto Nazionale di Fisica Nucleare, Italy; Ministry of Development of Turkey under Contract No. DPT2006K-120470; National Research Foundation of Korea under Contract No. NRF-2022R1A2C1092335; National Science and Technology fund; National Science Research and Innovation Fund (NSRF) via the Program Management Unit for Human Resources \& Institutional Development, Research and Innovation under Contract No. B16F640076; Polish National Science Centre under Contract No. 2019/35/O/ST2/02907; The Royal Society, UK under Contract No. DH160214; The Swedish Research Council; U. S. Department of Energy under Contract No. DE-FG02-05ER41374

\end{acknowledgments}

\end{document}